\documentclass[5p]{elsarticle}
\usepackage{booktabs,shortvrb}
\usepackage{amsmath,amsfonts,amssymb}%
\usepackage{booktabs,url}

\makeatletter
    \def\ps@pprintTitle{\let\@oddhead\@empty\let\@evenhead\@empty\let\@oddfoot\@empty\let\@evenfoot\@oddfoot }
\makeatother
\begin{document}
\begin{frontmatter}
\title{Multi-agents features on Android platforms}
\author[1]{Camelia-M. Pintea}\ead{dr.camelia.pintea@ieee.org}
\author[2]{Andreea Camelia Tripon}\ead{camelia.tripon@rodeapps.com} 
\author[1]{Anca Avram}\ead{anca.avram@ieee.org}
\author[3]{Gloria-Cerasela Crisan }\ead{cerasela.crisan@ub.ro}
\address[1]{Technical University Cluj Napoca, North University Center Baia Mare, Romania}
\address[2]{RodeApps, Cluj-Napoca, Romania}
\address[3]{"Vasile Alecsandri" University, Bac\u au, Romania}
\begin{abstract}
 The current paper shows the multi-agents capabilities to make a valid and flexible application when using a framework. Agent-based functions were used within JADE framework to make an Android messenger application with all requirements included. In the paper are described the architecture, the main functions and the databases integration of the user friendly agent-based application. There are included existing and possible multi-agents characteristics to provide integration with mobile platforms and  storage challenges to improve the user experience through data mining.
\end{abstract}
\end{frontmatter} 
\section{Introduction}
Multi-agent systems are involved today for solving different type of problems. They could be used in real-time applications and for solving complex problems in different domains as bio-informatics, ambient intelligence, semantic web~\cite{jsw,wm}. The main properties of agents in general, including the JADE~\cite{jade} platform used here, are the autonomy, reactivity, pro-activeness, cooperation, mobility and not at last learning capabilities.
 
JADE (Java Agent Development Framework) is a Java-based framework used to specify the multi-agent systems. A JADE-based system can be distributed over many systems and its configuration could be controlled for example by a user-friendly graphical application. 

The communication architecture of JADE offers an efficient and flexible transmission of messages based on a private queue of messages under {\it Agent Communication Language (ACL)} format for each agent. The agents can easily identify the ACL messages received from other agents and have access to their queue of messages. 

The Android applications are nowadays some of the most used application, especially due to the free operating system features. Known to be frequently used, the messenger applications are real-time communication means between people. The messenger could be installed and used through any existing Android devices as smartphones or tablets. 

The structure of the paper includes multi-agents with specific features for Android applications in the second section. In Section 3 the messenger Application based on JADE is introduced with the description of the structure and the main functionalities of the messenger application. Section 4 includes new approaches on agents features for Android applications including GPS/GIS characteristics. Section 5 illustrates the storage challenges and improving the user experience through data mining. The conclusion of the paper presents the new JADE-based messenger and possible multi-agent future improvements. 

\section{Multi-Agent Systems enhance the Android applications features}
In several disciplines as Artificial Intelligence, in human-computer interface design and in object-based systems~\cite{jsw} the intelligent agents have been used. The agents are inspired from real living "agents" as humans, insects or animals capable to react to their own environment, with own objectives to reach and the autonomous capability to make the proper activities to achieve its goals~\cite{gecco,ljigpl}.

In general, it is assumed that an agent has the following properties \cite{fg,wm}:
\begin{itemize}
\item The autonomy shows the ability of the agent to operate by itself without any other intervention.
\item The reactivity shows the ability of the agent to know the environment and react to the changes from its environment.
\item The pro-activeness shows the ability of the agent to have initiative and pursue its own goals.
\item The cooperation shows the capability of the agent to interact with others,  agents or humans, through a specific communication language.
\item The learning ability of the agent is activated while the agent interacts with its environment.
\item The mobility shows the ability of an agent to move in a self-directed way  around a network.  
\end{itemize}

Based on the particularities of a problem to be solved, the agents could be endowed with other features as for example rationality or sensitivity. The agents from the {\em Multi-Agent System (MAS)} are autonomous and heterogeneous agents capable of interaction. In MAS the computation is asynchronous and has no global control~\cite{jsw}.

In \cite{jsw} is specified that negotiation, as a coordination process, is essential in {\em MAS} to solve conflicts. The communication~\cite{jsw} in {\em MAS} is a must due to exchanging information or other inter-operation tasks between agents. The communication process between agents requires the {\it Agent Communication Language (ACL)} and understanding the concepts exchanged by agents.

The specific literature~\cite{poslad,httproma,httpdev} presents many connections between the evolution and the organization of a MAS and the way an Android applications works and is coded. These strong connections generated in fact the JADE project, its worldwide success both in academia and in industry.

\noindent Basically, the event-driven paradigm materialized by a generic Android application can be seen as the theoretical specification of the behaviour of MAS. In the following it is presented such an application that couples the MAS paradigm with Android.

\section{A new messenger application based on JADE}

An application based on JADE is made from a set of components called agents. The agents are uniquely identified by their names. The agents execute tasks and interact by changing messages. The agents are kept "alive" through a platform offering services. A platform has one or many containers that could be run on systems with different hosts. Each container could have zero or many agents. Each platform has a special container called the  Main Container with different agents (Figure 1).
 \begin{figure}[ht]\label{fig:1}\centering \includegraphics[scale=0.35]{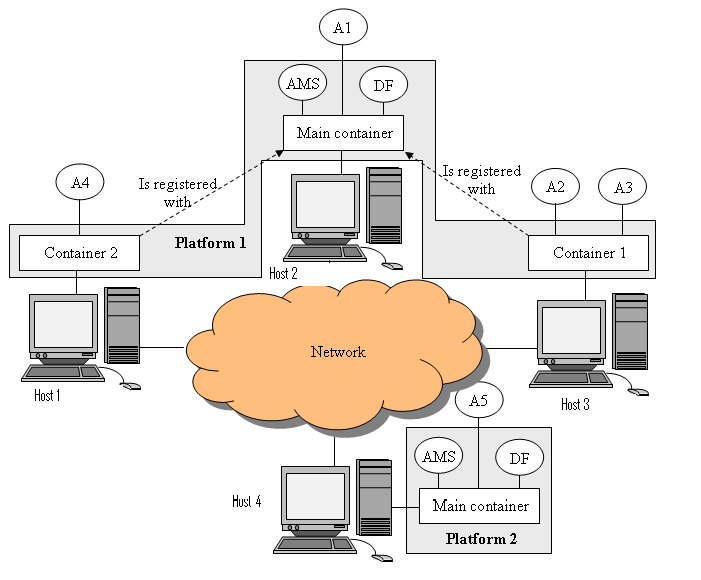}\caption{The main architecture of JADE framework~\cite{jade}.}                        \end{figure}  

The minimal installation and running requirements are: minimum JDK 5, Android Studio and  Android SDK. Once the Main Container is started, the server starts too. The other agents will run from the same or different system and will connect with the Main Container. After that, the application is ready to function as a real-time messenger application.

When an agent is connected to the Main Container, a new container will be created and will be visible; to run other agents we can install the messenger application on an Android phone, run the application and then connect to the Main Container.
   
The Main Container is the container that starts at first and the other containers will login to it; the Main Container includes two special agents: 
\begin{itemize}
\item The {\it Agent Management System (AMS)} is the authority of the platform and it is the only agent that could manage the platform (starting or stopping  agents or stopping the entire platform).
\item The {\it Direction Facilities} (DF) is a service through which the agents could publish the tasks offered and give the possibility to find other agents based on the task.
\end{itemize}

The communication between agents is made no matter if the agents are or not in the same container or if they belong or not to the same platform. The messages format ACL is defined by {\it Foundation for Intelligent Physical Agents (FIPA)}. An ACL message includes the sender, the receiver,  the communication task and the content of the message. There are twenty two communication FIPA sets, each one well defined and having a well defined semantic. For example INFORM message: "INFORM 7 May 2016 rains", or REQUEST message requesting the receiver to make a task.

\subsection{The messenger application}

Multi-agent system is used for the application, a JADE-based messenger on the Android operating system. A device (phone/tablet) with Android operating system connects with the server address; after the connection is established, the agent could send and receive messages. 

The application considers the client-server principle: through the already existing connections, the clients ask the server to do some tasks like sending messages. When the server receives the message, the server sends it further to all the connected clients. The application has two modules:

\begin{itemize}
\item[1.] The Android Messenger Client permits the messenger connection from an Android device; it includes a graphical interface and an agent managing the interactions with the other components.

\item[2.] The Messenger Server is the platform called Main Container including an agent called {\it ChatManagerAgent} keeping the evidence of all agents connected to the messenger.

\end{itemize}

\noindent The steps to connect to the Main Container are the following:
 \begin{itemize}
 \item The installation of application on the Android phone or on a simulator. 
 \item The IP address of the server is set from the menu and should be the same with the address of the server.
      
\item User must choose a name, which should be different from the name of the other messenger-users.
\item If there are no errors, the messenger-user could send messages to the others messenger-users. 
\end{itemize}

\subsection{The structure of the messenger}

The application has two parts. The first part makes the connection with the JADE agent and the second one is the graphical interface. They are included in the packages \texttt{agent} and  \texttt{gui}. Based on the Android architecture, the JADE agent is integrated in the project through {\it jade-Android.jar} library; it includes the following services to connect to the server:
{\it RuntimeService} and {\it MicroRuntimeService}.

In this particular application it is used the {\it MicroRuntimeService}. The connectivity with the service is made by the class {\it MainActivity} through the following code:

\begin{small}
\begin{verbatim}

serviceConnection = new ServiceConnection(){
 public void onServiceConnected(ComponentName name, 
                                IBinder service){
  microRuntimeServiceBinder = (MicroRuntimeServiceBinder) 
	                             service;
 };
 public void onServiceDisconnected(ComponentName name){
  microRuntimeServiceBinder = null;
 }
};

bindService(new Intent(getApplicationContext(),
 MicroRuntimeService.class),serviceConnection, 
 Context.BIND_AUTO_CREATE);

\end{verbatim}
\end{small}
Once connected to the service, we can go to the next step, which is to create the container and to start the agent. If it is a success, then the application can send and receive messages. As a JADE-messenger novelty, there are included many actions further described including for example the groups of users, the messages with a single user, the messages within a group, blocking/unblocking users, etc.

\begin{small}
\begin{verbatim}
    void getAllUsers();
    void getUsersNotInGroup(Group group);
    void getBlockedUsers();
    void getGroupsConversations();
    void getUsersFromGroup(Group group);
    void getMessagesFromUser(User user);
    void getMessagesFromGroup(Group group);
    void getUserConversations();
    void createGroup(Group group);
    void leaveGroup(Group group);
    void addToGroup(User user, Group group);
    void blockUser(User user);
    void unblockUser(User user);
    void register(User user);
    void login(User user);
    void sendMessageToGroup(Message message);
    void sendMessageToUser(Message message);
    void deleteMessage(Message message);
    ArrayList<User> getParticipantNames();
\end{verbatim}
\end{small}
The first method is called when a message is sent from the phone and the second when the user wants to see the other messenger-users connected to the same server.
These two actions are on the \texttt{ChatClientAgent} class, implementing the \texttt{ChatClientInterface}. This class deals with sending and receiving the messages further to/from the \texttt{ChatActivity} class in order to visualize the messages. In the \texttt{ChatClientAgent} class it is implemented a listener; when the server send a message, the message is shown through a broadcast receiver.

\subsection{The main functions of the application}
The introduced application is a stand-alone messenger application for Android with the main functions of any Google Play Store application with a user-friendly design. The application starts from the main menu, after the user logged in or registered in the application. If some of the settings are not correct, the user could change the port or the host address. To set the connection the address where the JADE agent would be connected, the host and the port should be specified. Implicitly are used the last ones used for a connection. 
 
Logging into the system is beneficent to keep all the user information saved and the user could use different mobile phone to send/receive the messages, or see the older messages. 
 
The button used to show all the users will open a page with the status of all  registered users. The current user could add any user of an existing group or block another user. If a user is blocked, he does not have the possibility to send messages to the one who blocked him. In the same context, you cannot see the status of an blocked user and cannot add him to a group. These could be possible after the user is unblocked (Figure 2).
 
\begin{figure}\centering \includegraphics[scale=0.4]{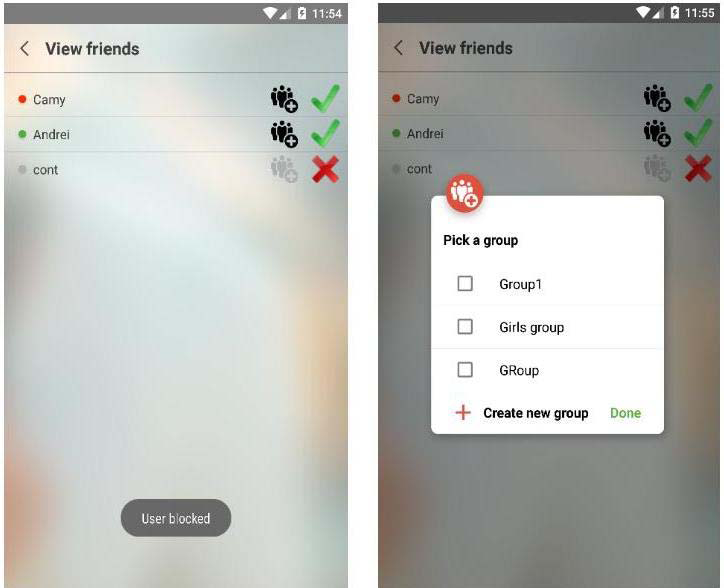} \caption{Examples of the messenger-JADE features: blocking/unblocking users (left) and managing groups of users (right).}\end{figure}
  
A new group is created easily using a particular button and by specifying its name. The name is verified to be unique in the database and an appropriate message is shown. The new group is automatically added to the other existing groups. Nevertheless, the only ones who can add other members to a group are the members of the groups.
 
When selecting a conversation with a friend all the sent and
received messages from that user will be shown. On the toolbar there are the user-name, a button to add the friend to the group, a button to add the user in the blocking list and a button to delete all the conversation. In the left side of the window there are the received messages and in the right side the sent messages. All the messages include the time and date when they were sent. At the bottom of the window are the buttons for adding and sending new messages. To delete a message, one can use a long touch on that message and the message could be deleted after a confirmation.
 
For the conversations in a group there is a similar functionality; the differences are on the toolbar: adding a new member to the group, visualization of all the group members and their status, the button to leave the group. All the messages received from any member of the group will be placed in the left side of the window and the sent ones in the right side of the window.                                                                                   
\section{Theoretic approach on new agents features for Android applications}

Several common features for Multi Agent Systems and an Android application are:
\begin{itemize}
\item Concurrency: the Android components are activated by intents (decentralized events trigger an agent behavior or the Yellow Pages service provided by the DF special JADE agent~\cite{jade}). 

The agents from MAS act simultaneously, in a common environment, pursuing their individual goals, but possibly ready to draw coalitions or to co-operate.
\item Loosed coupling: like agents in MAS, the Android application target components are activated by explicit or implicit intents~\cite{poslad}.
\item Asynchronous communication: both for the agents in MAs and for the Android application components, the self-decided conversation initiation is a manifestation of the individual, autonomous behavior.

\item Flexibility: Android is a multi-channel, multi-carrier, freely distributed OS, based on Java, and has a market share of 82.5\% in 2015~\cite{httpidc}. 

The implicit intent object is an example of how the Android operating system adapts to the environment~\cite{httpdev}. Likewise, current complex social problems are modeled by MAS's that need to perform well in open environments. 

For example, the MIT Robust Open Multi-Agent Systems (ROMA) Research Group is dedicated to "…learning how we can develop multi-agent systems for open contexts where the constituent agents can come from anywhere, may be buggy or even malicious, and must run in the dynamic and potentially failure-prone environments at hand"~\cite{httproma}.
\end{itemize}

The agents involved in the JADE messenger application have limited features. Here we introduce several new features to be included in the JADE framework.

The Multi-agent system of JADE could include agents with different levels of sensitivity related to their environment. Based on the agents sensitivity, the messenger application could include other facilities as:
\begin{itemize}
\item the messenger could start when the user is closer to a already set point; for example one could set the messenger application to start when approaching to a building, let's say a museum, when the application starts automatically based on the GPS/GIS; so, the agent is sensitive to each "museum" encountered to exchange impressions with the other users.

\item the messenger could include a blocking option based on the agent sensitivity: an user is automatically blocked or unblocked based on its behavior:

\begin{itemize}
\item if the words used by another user are "bad", so they are in a database, locally or cloud-based, the user is automatically blocked and an appropriate message is sent to the other user;

\item an user could be unblocked if it is identified a good behavior on its personal social platform; for example on its social website there are included voluntary actions, promoting "good words" and facts.  This feature could be added or not by the user.

\end{itemize}

\item another improvement could be considered the automatically connection to a close or extended group when an earthquake is ongoing based on a specific sensor of the device or by an earthquake alert. Similar features could be included for floods or other related crisis. 

The GPS/GIS features could be used as in~\cite{moratis,cai,lei,crisan}. In~\cite{cai} is shown an implementation of an intelligent, multi-modal, multi-user geographic information environment (GCCM\_Connect) used on a spatial decision-making contexts. Collaboration and share knowledge is a must especially when is involved a critical problem as for example earthquakes or floods. Crisis management demands information technology and individuals /organizations to share information and expertise on decision-making.  

\item An idea of improvement in {\bf terms of functionality} would be to have the existing list of named servers available and user to choose from the server list instead of having to configure the server himself the host and port. When blocking another user, one could invoke a reason for doing that if he/she wants. There could be a list of predefined reasons or user can invoke a new one. Later on, analysis of data could show the potential users that have malicious intents. 
Auto-complete features could be provided based on analyzing the most used phrases and offer edit support to the users using the application.

\end{itemize}
All the new and the common characteristics generate tight connections, allowing the researchers to design and to provide useful integration of the mobile platform with Multi Agent System features. 

\section{Storage challenges and improving the user experience through data mining}

For the messenger application presented, there was the need of having information stored in a database. The information that is stored consists of details related to users, login information, friends of a user, blocked/unblocked users; messenger groups; messages exchanged between users; agent behaviors.

As soon as the application is used by more and more users, the volume of the data increases and a proper database maintenance plan needs to be in place. That means there will be a need for performing actions like:
\begin{itemize}
\item	Cleaning unnecessary data - logs of the status of actions performed by the agents for example, are only needed for a short amount of time (in case of an error for example, these logs could provide a meaningful support for finding the cause of the problem).

\item	Archiving the old data. A proper system will make sure that data that is not in use anymore is handled (for example messages exchanged in groups that no longer contain active users). Archiving could mean storing the plain data to a different server (less performing, less expensive). Another approach would be obtaining a more compact version of that data and storing it in a different format.

\item	Having an indexing strategy that is reviewed periodically. Indexes are data structures that improve the speed of data retrieval operations on a table. Having proper indexes will mean a faster time to obtain the results, but reviewing the indexes periodically is a must, because the increase of data could mean that an index that was performing at some point might no longer be efficient.
\end{itemize}

The database could also provide the means for obtaining relevant information related to the behavior of the application and agents by performing data mining. Currently data mining is widely recognized as the process of discovering relevant patterns in large sets of data, patterns that can be later used. To have these patterns discovered, intelligent methods are applied. Data mining can be applied to any kind of data, as long as it delivers meaningful results to a target application~\cite{han2011}.

In the context of the JADE messenger application, data mining could be applied on the stored data related to the users to obtain relevant information on the user behavior. For example, discovering the use patterns for a specific functionality could give an idea on what type of users are using that functionality, what are the topics that are the most tackled, what is the geographical distribution of the users, whether the application is more widely used in the urban or rural area. This type of knowledge could lead further on to developing new functions that are targeted to improving user experience; remove/make less visible functions that do not come with an adequate return of investment.

Furthermore, in~\cite{cao2009} the concept of agent mining refers to the application of autonomous intelligent agents in the field of data mining to support and improve the knowledge discovery and decision-making process. Because agents have autonomy, flexibility, mobility, adaptability and have a rational nature, they prove to be a perfect choice for parallel, multisource, distributed mining. In the context of the JADE framework, that could mean integrating agents responsible with mining the data.

\section{Conclusions and future work}
The paper introduces a multi-agent-based messenger application using JADE framework. JADE has several components successfully used for the current messenger application as the {\it Agent Communication Language (ACL)} to send/receive the messages. The user-friendly application includes several features as using groups of users or blocking/unblocking users on any Android devices. The application could be improved with for example audio/video messages and especially with other multi-agents features, as sensitivity based on GPS/GIS location.
 
\smallskip
 
\noindent{\bf Acknowledgments.} The study was conducted under the auspices of the IEEE-CIS Interdisciplinary Emergent Technologies TF. We thank Alexandru Pintea for carrying out Latex editing tasks.

\end{document}